\documentclass{llncs}

\usepackage{epsfig}
\usepackage{psfig}
\long\def\comment#1{}
\newcommand{\cT}{\mbox{${\cal T}$}}
\newcommand{\cP}{\mbox{${\cal P}$}}
\newcommand{\compact}{\vspace{-0.4em}}

\newcommand{\tA}{\mbox{$\texttt{A}$}}
\newcommand{\tC}{\mbox{$\texttt{C}$}}
\newcommand{\tT}{\mbox{$\texttt{T}$}}
\newcommand{\tG}{\mbox{$\texttt{G}$}}
\newcommand{\tS}{\mbox{$\texttt{S}$}}
\newcommand{\tW}{\mbox{$\texttt{W}$}}

\newcommand{\WciiiS}{\mbox{$\tW$$<$$c-3$$>$$\tS$}}
\newcommand{\ScivS}{\mbox{$\tS$$<$$c-4$$>$$\tS$}}
\newcommand{\SciiiS}{\mbox{$\tS$$<$$c-3$$>$$\tS$}}   
\newcommand{\WciiW}{\mbox{$\tW$$<$$c-2$$>$$\tW$}}  
\newcommand{\SciiiW}{\mbox{$\tS$$<$$c-3$$>$$\tW$}}
\newcommand{\SciiiWW}{\mbox{$\tS$$<$$c-3$$>$$\tW\tW$}}
\newcommand{\ScivSW}{\mbox{$\tS$$<$$c-4$$>$$\tS\tW$}}

\newcommand{\sWciiiS}{\mbox{$\scriptsize{\WciiiS}$}}
\newcommand{\sScivS}{\mbox{$\scriptsize{\ScivS}$}}
\newcommand{\sSciiiS}{\mbox{$\scriptsize{\SciiiS}$}}   
\newcommand{\sWciiW}{\mbox{$\scriptsize{\WciiW}$}}  
\newcommand{\sSciiiW}{\mbox{$\scriptsize{\SciiiW}$}}
\newcommand{\sSciiiWW}{\mbox{$\scriptsize{\SciiiWW}$}}
\newcommand{\sScivSW}{\mbox{$\scriptsize{\ScivSW}$}}

\begin{document}

\title{Improved Tag Set Design and\\ Multiplexing Algorithms 
for Universal Arrays\thanks{Work supported in part by a 
``Large Grant'' from the University of Connecticut's 
Research Foundation.}}

\author{Ion I. M\u{a}ndoiu
\and 
Claudia Pr\u{a}jescu
\and
Drago\c{s} Trinc\u{a}
}

\institute{CSE Department, University of Connecticut\\
371 Fairfield Rd., Unit 2155, Storrs, CT 06269-2155\\
\email{\{ion.mandoiu,claudia.prajescu,dragos.trinca\}@uconn.edu} 
}

\maketitle

%%%%%%%%%%%%%%%%%%%%%%%%%%%%%%%%%%%%%%%%%%%%%%%%%%%%%%%%%%%%%%%%%%

\begin{abstract}
In this paper we address two optimization problems arising in the
design of genomic assays based on universal tag arrays.
First, we address the universal array tag set design problem. 
For this problem, we extend previous formulations to incorporate 
antitag-to-antitag hybridization constraints in addition to 
constraints on antitag-to-tag hybridization specificity, establish 
a constructive upper bound on the maximum number of tags satisfying 
the extended constraints, 
and propose a simple greedy tag selection algorithm.
Second, we give methods for improving the multiplexing rate               
in large-scale genomic assays 
by combining primer selection with tag assignment.
Experimental results on simulated data show that this integrated optimization 
leads to reductions of up to 50\% in the number of required arrays.
\end{abstract}

%%%%%%%%%%%%%%%%%%%%%%%%%%%%%%%%%%%%%%%%%%%%%%%%%%%%%%%%%%
\section{Introduction}
\label{sec.intro}

High throughput genomic technologies have revolutionized biomedical 
sciences, and progress in this area continues at an accelerated pace
in response to the increasingly varied needs of biomedical research.
Among emerging technologies, one of the most promising is 
the use of {\em universal tag arrays} \cite{Brenner97,affy_patent,Gerry99},
which provide unprecedented assay customization flexibility while 
maintaining a high degree of multiplexing and low unit cost.

A universal tag array consists of a 
set of DNA {\em tags}, designed such that each 
tag hybridizes strongly to its own {\em antitag} (Watson-Crick complement), 
but to no other antitag \cite{Karp00}.
Genomic assays based on universal arrays involve 
multiple hybridization steps.
A typical assay \cite{BenDor03,Hirschhorn2000}, 
used for Single Nucleotide Polymorphism (SNP) genotyping, works as follows.
(1) A set of {\em reporter oligonucleotide probes} 
is synthesized by ligating antitags to the $5'$ end of primers complementing 
the genomic sequence immediately preceding the SNP location in $3'$-$5'$ 
order on either the forward or reverse strands.
(2) Reporter probes are hybridized in solution with the genomic 
DNA under study. 
(3) Hybridization of the primer part ($3'$ end) of a reporter probe
is detected by a single-base extension reaction using
the polymerase enzyme and dideoxynucleotides fluorescently labeled 
with 4 different dyes. 
(4) Reporter probes are separated from the template DNA 
and hybridized to the universal array.  
(5) Finally, fluorescence levels 
are used to determine which primers have been extended and learn 
the identity of the extending dideoxynucleotides.

In this paper we address two optimization problems arising in the 
design of genomic assays based on the universal tag arrays.
First, we address the universal array {\em tag set design problem}
(Section \ref{sec.tag-set-design}). To enable the economies 
of scale afforded by high-volume production of the arrays, 
tag sets must be designed to work well for a wide range of assay types and 
experimental conditions. Ben Dor et al. \cite{Karp00} have 
previously formalized the problem by imposing constraints on 
antitag-to-tag hybridization specificity under a hybridization 
model based on the classical 2-4 rule \cite{Wallace}. We extend the model in 
\cite{Karp00} to also prevent antitag-to-antitag hybridization 
and the formation of antitag secondary structures, 
which can significantly interfere with 
or disrupt correct assay functionality. Our results on this problem 
include a constructive upper bound on the maximum number of tags 
satisfying the extended constraints, as well as a simple 
greedy tag selection algorithm.

Second, we study methods for improving the multiplexing rate 
(defined as the average number of reactions assayed per 
array) in large-scale genomic assays involving multiple universal arrays.  
In general, it is not possible to assign all tags to primers in an array 
experiment due to, e.g., unwanted primer-to-tag hybridizations. 
An assay specific optimization that determines the multiplexing rate
(and hence the number of required arrays for a large assay) 
is the {\em tag assignment problem}, 
whereby individual (anti)tags are assigned to each primer. 
In Section \ref{sec.integrated} we observe that 
significant improvements in multiplexing rate can be achieved 
by combining primer selection with tag assignment.  
For most universal array applications there are multiple primers with 
the desired functionality; for example in the SNP genotyping assay 
described above one can choose the primer from either 
the forward or reverse strands.  Since different primers 
hybridize to different sets of tags, a higher multiplexing rate is 
achieved by integrating primer selection with 
tag assignment. This integrated optimization
\comment{, inspired from 
the flow enhancement techniques developed in the 
field of electronic design automation \cite{IBM,Sherwani},}
is shown in Section \ref{sec.results} 
to lead to a reduction of up to 50\% in the number of required arrays.

%%%%%%%%%%%%%%%%%%%%%%%%%%%%%%%%%%%%%%%%%%%%%%%%%%%%%%%%%%
\section{Universal Array Tag Set Design} 
\label{sec.tag-set-design}

The main objective of universal array tag set design 
is to maximize the number of tags, 
which directly determines the number of reactions that 
can be multiplexed using a single array. 
Tags are typically required to have a predetermined length 
\cite{affy_note,affy_patent}.  Furthermore, for 
correct assay functionality, tags and their 
antitags must satisfy the following hybridization 
constraints:
\begin{description}
\compact
\item 
(H1) Every antitag hybridizes strongly to its tag;
\item 
(H2) No antitag hybridizes
to a tag other than its complement; and 
\item
(H3) There is no antitag-to-antitag hybridization (including 
hybridization between two copies of 
the same tag and self-hybridization), since the formation of such 
duplexes and hair-pin structures prevents corresponding reporter probes from 
hybridizing to the template DNA and/or leads to 
undesired primer mis-extensions. 
\compact
\end{description}

Hybridization affinity between two oligonucleotides is 
commonly characterized using the {\em melting temperature}, 
defined as the temperature at which 
50\% of the duplexes are in hybridized state.  
As in previous works \cite{Karp00,BenDor03}, 
we adopt a simple hybridization model to formalize constraints (H1)-(H3).
This model is based on the 
observation that stable hybridization requires the 
formation of an initial {\em nucleation complex}
between two perfectly complementary 
substrings of the two oligonucleotides.
For such complexes, hybridization affinity is well  
approximated using the classical {\em 2-4 rule} 
\cite{Wallace}, which estimates the melting temperature 
of the duplex formed by an oligonucleotide with  
its complement as the sum between twice the number of $\tA$+$\tT$ bases
and four times the number of $\tG$+$\tC$ bases. 
\comment{Based on these assumptions, we formalize the tag set design 
problem as follows.}

The {\em complement} of a string $x=a_1a_2\ldots a_k$ 
over the DNA alphabet $\{\tA,\tC,\tT,\tG\}$ is 
$\bar{x}=b_1b_2\ldots b_k$, where $b_i$ is the 
Watson-Crick complement of $a_{k-i+1}$.
The {\em weight} $w(x)$ of $x$ is defined as
$w(x) = \sum_{i=1}^{k}w(a_i)$, where 
$w(\tA)=w(\tT)=1$ and 
$w(\tC)=w(\tG)=2$.

\begin{definition} \label{def.feasible-tag-set}
For given constants $l$, $h$, and $c$ with 
$l\le h\le 2l$, 
a set of tags $\cT\subseteq \{\textup{\tA},\textup{\tC},\textup{\tT},\textup{\tG}\}^l$ is called 
{\em feasible} if the following conditions 
are satisfied:
\begin{itemize}
\compact
\item \mbox{(C1)} Every tag in $\cT$ has weight $h$ or more.
\item \mbox{(C2)} Every DNA string of weight $c$ or more 
appears as substring at most once in the tags of $\cT$.
\item \mbox{(C3)} If a DNA string $x$ of weight $c$ or more 
appears as a substring of a tag, 
then $\bar{x}$ does not appear as a substring of a tag 
unless $x=\bar{x}$.
\compact
\end{itemize}
\end{definition}

The constants $l$, $h$, and $c$ depend on factors such as array
manufacturing technology and intended hybridization conditions.
Property (H1) is implied by (C1) when $h$ 
is large enough. Similarly, properties (H2) and (H3) are 
implied by (C1) and (C2) when $c$ is small enough: 
constraint (C2) ensures that nucleation complexes do not 
form between antitags and non-complementary tags,
while constraint (C3) ensures that nucleation complexes 
do not form between pairs of antitags.

\smallskip
\noindent
{\bf Universal Array Tag Set Design Problem:}  
{\em Given constants $l$, $h$, and $c$ with
$l\le h\le 2l$, find a feasible tag set of maximum cardinality.}
\smallskip

Ben-Dor et al. \cite{Karp00} have recently studied 
a simpler formulation of the problem in 
which tags of unequal length are allowed and 
only constraints (C1) and (C2) are enforced.  
For this simpler formulation, Ben-Dor et al. established a 
constructive upperbound on the optimal number of tags, 
and gave a nearly optimal tag selection algorithm based 
on De Bruijn sequences.  
Here, we refine the techniques in \cite{Karp00} to establish 
a constructive upperbound on the number of tags of a feasible 
set for the extended problem formulation, and propose a 
simple greedy algorithm for constructing feasible tag sets.

The constructive upperbound is based on counting the minimal 
strings, called {\em $c$-tokens}, that can occur as 
substrings only once in the tags and antitags of a feasible set. 
Formally, a DNA string $x$ is called $c$-token if the weight of $x$ 
is $c$ or more, and every proper suffix of $x$ has weight 
strictly less than $c$.  
The {\em tail weight} of a $c$-token 
is defined as the weight of its last letter. 
Note that the weight of a $c$-token 
can be either $c$ or $c+1$, the latter case being possible only 
if the $c$-token starts with a $\tG$ or a $\tC$.  
As in  \cite{Karp00}, we use $G_n$ to denote the number of 
DNA strings of weight $n$. It is easy to see that 
$G_1=2$, $G_2=6$, and $G_n=2G_{n-1} + 2G_{n-2}$; for convenience, 
we also define $G_0=1$. In Appendix \ref{sec.appendix} we prove the following:

\begin{lemma} \label{lemma.count}
Let $c\ge 4$. 
Then the total number of $c$-tokens that appear as substrings 
in a feasible tag set is at most 
$3G_{c-2} + 6G_{c-3} + G_{\frac{c-3}{2}}$ 
if $c$ is odd, and at most 
$3G_{c-2} + 6G_{c-3} + \frac{1}{2}G_{\frac{c}{2}}$ 
if $c$ is even. 
Furthermore, the total tail weight of 
$c$-tokens that appear as substrings 
in a feasible tag set is at most 
$2G_{c-1}+4G_{c-3}+2G_{\frac{c-3}{2}}$
if $c$ is odd, and at most
$2G_{c-1}+4G_{c-3}+G_{\frac{c-2}{2}}+2G_{\frac{c-4}{2}}$
if $c$ is even.
\end{lemma}

\begin{theorem} \label{thm.upperb}
For every $l$, $h$, $c$ with $l\le h\le 2l$ and $c\ge 4$,   
the number of tags in a feasible tag set is at most 
\compact
\[
\min
 \left\{
   \frac{ 3G_{c-2} + 6G_{c-3} + G_{\frac{c-3}{2}} }{l-c+1},
   \frac{ 2G_{c-1}+4G_{c-3}+2G_{\frac{c-3}{2}} }{h-c+1}
 \right\}
\compact
\compact
\]
for $c$ odd, and at most 
\compact
\[
\min
 \left\{
    \frac{ 3G_{c-2} + 6G_{c-3} + \frac{1}{2}G_{\frac{c}{2}} }{l-c+1},
    \frac{ 2G_{c-1}+4G_{c-3}+G_{\frac{c-2}{2}}+2G_{\frac{c-4}{2}} }{h-c+1}
 \right\}
\compact
\compact
\]
\compact
for $c$ even.
\end{theorem}

\begin{proof}
The proof follows from Lemma \ref{lemma.count} by observing that 
every tag contains at least $l-c+1$ $c$-tokens, with a total 
tail weight of at least $h-c+1$.
\qed
\end{proof}

We employ a simple greedy algorithm to generate feasible 
sets of tags; a similar algorithm is suggested in \cite{affy_patent} 
for finding sets of tags that satisfy an unweighted version of 
constraint (C2). 
We start with an empty set of tags and 
an empty tag prefix. In every step we try to extend the 
current tag prefix $t$ by an additional $\tA$.  
If the added letter completes a $c$-token or a complement 
of a $c$-token that has been used in already 
selected tags or in $t$ itself, we try the next letter in the DNA alphabet, 
or backtrack to a previous position in the prefix when no more 
letter choices are left.  Whenever we succeed generating 
a complete tag, we save it and backtrack to the last letter of its 
first $c$-token.

%%%%%%%%%%%%%%%%%%%%%%%%%%%%%%%%%%%%%%%%%%%%%%%%%%%%%%%%%%
\section{Improved Multiplexing by Integrated Primer Selection and Tag Assignment} 
\label{sec.integrated}

\begin{figure}[t]                 
\compact
\centerline{\psfig{figure=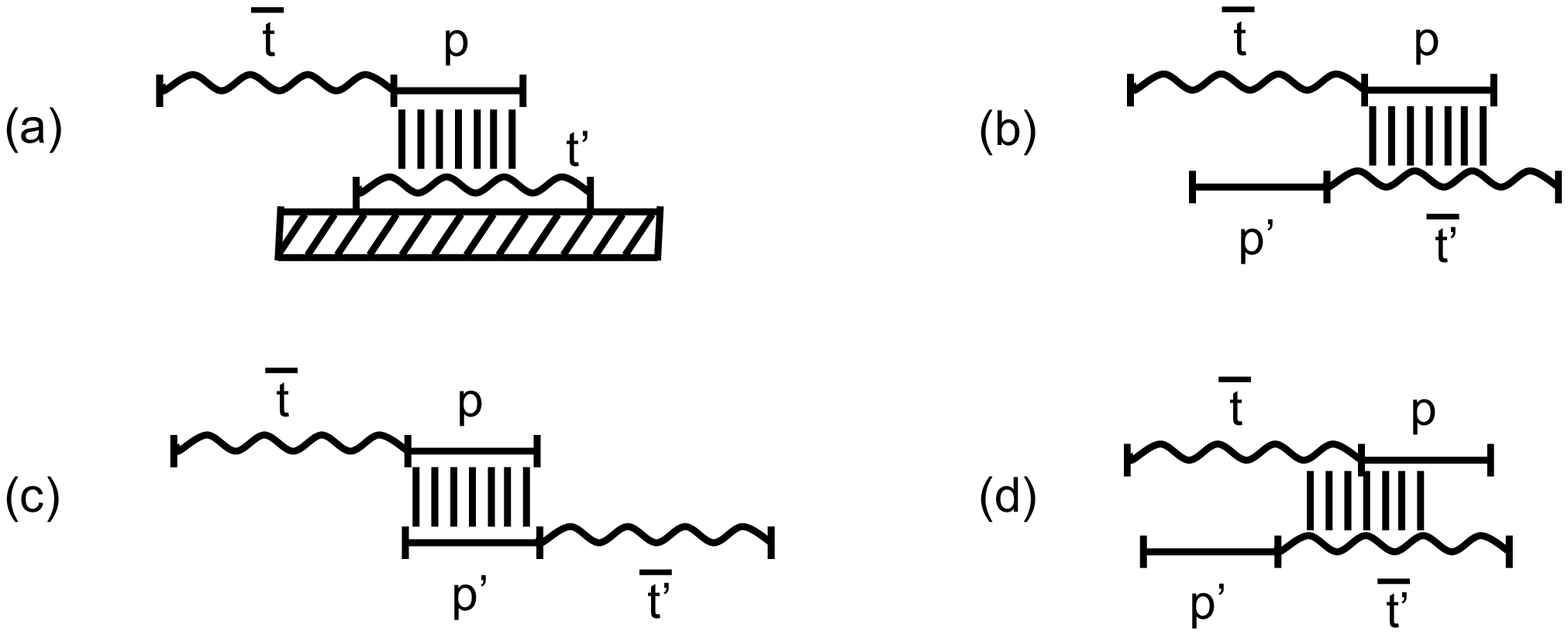,width=3in}}
\compact
\compact
\compact
\caption{\label{fig.crosspriming}Four types of undesired 
hybridizations, caused by the formation of nucleation complexes 
between 
(a) a primer and a tag other than the complement of the ligated antitag,
(b) a primer and an antitag, 
(c) two primers,
and 
(d) two reporter probe substrings, at least one of which 
straddles a ligation point.
}
\compact
\compact
\compact
\end{figure}  

Although constraints (H2)-(H3) 
in Section \ref{sec.tag-set-design} prevent 
unintended antitag-to-tag and antitag-to-antitag hybridizations,
the formation of nucleation complexes involving (portions of) the primers 
may still lead to undesired hybridization 
between reporter probes and tags on the array (Figure \ref{fig.crosspriming}(a)), 
or between two reporter probes (Figure \ref{fig.crosspriming}(b)-(d)).
The formation of these duplexes must be avoided as it 
leads to extension misreporting, 
false primer extensions, 
and/or reduced effective 
reporter probe concentration available for hybridization 
to the template DNA or to the tags on the array \cite{BenDor03}. 
This can be done by leaving some of the tags unassigned.
As in \cite{BenDor03}, we focus on preventing primer-to-tag hybridizations
(Figure \ref{fig.crosspriming}(a)). Our algorithms can be easily 
extended to prevent primer-to-antitag hybridizations (Figure \ref{fig.crosspriming}(b)); 
a simple practical solution for preventing the other (less-frequent)
unwanted hybridizations is to re-assign offending primers 
in a post-processing step.

Following \cite{BenDor03}, a set $\cP$ of primers is called 
{\em assignable} to a set $\cT$ of tags if there is a one-to-one 
mapping $a:\cP\rightarrow \cT$ such that, for every tag $t$ 
hybridizing to a primer $p\in\cP$, either $t\not\in a(\cP)$ or $t=a(p)$. 

\smallskip
\noindent
{\bf Universal Array Multiplexing Problem:} 
{\em Given primers  $\cP=\{p_1,\ldots,p_m\}$ 
and tag set $\cT=\{t_1,\ldots,t_n\}$, find a 
partition of $\cP$ into the minimum number of assignable sets.
}
\smallskip

For most universal array applications there are multiple primers with 
the desired functionality, e.g., for the SNP genotyping assay 
described in Section \ref{sec.intro}, one can choose the primer from either 
the forward or reverse strands. Since different primers        
have different hybridization patterns, a higher multiplexing rate can
in general be achieved by integrating primer 
selection with tag assignment. 
A similar integration has been recently proposed in \cite{iccd03}
between probe selection and physical DNA array design,
with the objective of minimizing unintended illumination in 
photo-lithographic manufacturing of DNA arrays.
The idea in \cite{iccd03} is to modify probe selection tools 
to return {\em pools} containing all feasible candidates, 
and let subsequent optimization 
steps select the candidate to be used from each pool. 
In this paper we use a similar approach. 
We say that a set of primer pools is {\em assignable} if 
we can select a primer from each pool to form an 
assignable set of primers.

\smallskip
\noindent
{\bf Pooled Universal Array Multiplexing Problem:} 
{\em Given primer pools $\cP=\{P_1,\ldots,P_m\}$ 
and tag set $\cT=\{t_1,\ldots,t_n\}$, find a 
partition of $\cP$ into the minimum number of 
assignable sets.
}
\smallskip

\begin{figure}[t]                 
\compact
\centerline{\psfig{figure=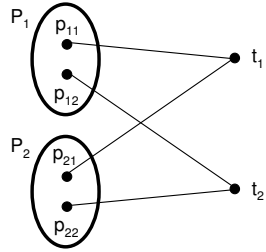,width=1.5in}}
\compact
\compact
\compact
\caption{\label{counter.fig}Two assignable pools for which 
 $|X(\cP)| + |Y(\cP)| = 0$.
}
\compact
%\compact
%\compact
\end{figure}

Let $\cP$ be a set of primer pools and $\cT$ a tag set. 
For a primer $p$ (tag $t$), $\cT(p)$ (resp. $\cP(t)$) 
denotes the set of tags (resp. primers of $\bigcup_{P\in \cP}P$) 
hybridizing with $p$ (resp. $t$). Let 
$
 X(\cP) = \{ P\in \cP ~:~ \exists p\in P,~t\in\cT \mbox{ s.t. } 
              t\in \cT(p) \mbox{ and } \cP(t)\subseteq P \}
$ 
and 
$
 Y(\cP) = \{ t\in \cT ~:~ \cP(t)=\emptyset \}
$.  
Clearly, in every pool of $X(\cP)$ we can find a primer $p$ that 
hybridizes to a tag $t$ which is not cross-hybridizing 
to primers in other pools, and therefore assigning $t$ to $p$ 
will not violate (A1).  Furthermore, any primer can be assigned 
to a tag in $Y(\cP)$ without violating (A1). 
Thus, a set $\cP$ with $|X(\cP)| + |Y(\cP)| \ge |\cP|$ is always 
assignable.  The converse is not necessarily true: 
Figure \ref{counter.fig} shows two pools that 
are assignable although $|X(\cP)| + |Y(\cP)| = 0$.

Our primer pool assignment algorithm (see Figure \ref{fig.primer-del})
is a generalization to primer pools of Algorithm B in \cite{BenDor03}.
In each iteration, the algorithm checks whether 
$|X(\cP')| + |Y(\cP')| \ge |\cP'|$ for the remaining 
set of pools $\cP'$. If not, a primer of maximum {\em potential} 
is deleted from the pools.  As in \cite{BenDor03}, the potential 
of a tag $t$ with respect to $\cP'$ is 
$2^{-|\cP'(t)|}$, and the potential of a primer $p$ is 
the sum of potentials for the tags in $\cT(p)$.
If the algorithm deletes the last primer in a pool $P$, then 
$P$ is itself deleted from $\cP'$; deleted pools are subsequently 
assigned to new arrays using the same algorithm.

\begin{figure}[t]             
\compact
{\footnotesize
\fbox{
\begin{minipage}{\textwidth}             
\begin{tabbing}
\hspace*{5mm}\=\hspace{5mm}\=\hspace{5mm}\=\hspace{5mm}\=\hspace{5mm}\=\hspace{5mm}\=  \kill
{\tt Input:} Primer pools $\cP=\{P_1,\dots,P_m\}$ and tag set $\cT$\\
{\tt Output:} Triples $(p_i,t_i,k_i)$, $1\leq i\leq m$, where
 $p_i\in P_i$ is the selected primer for pool $i$,\\
$t_i$ is the tag assigned to $p_i$, and $k_i$ is the index of the array on which $p_i$ is assayed \\
\rule[3pt]{1.0\textwidth}{0.3pt}\\
$k \leftarrow 0$ \\
{\tt While} $\cP \neq \emptyset$ {\tt do}\\
\>\>  $k \leftarrow k+1$;~  $\cP' \leftarrow \cP$\\
\>\>  {\tt While} $|X(\cP')| + |Y(\cP')| < |\cP'|$ {\tt do}\\ 
\>\>\> Remove the primer $p$ of maximum potential from the pools in $\cP'$\\
\>\>\> {\tt If} $p$'s pool becomes empty {\tt then} remove it from $\cP'$\\
\>\>{\tt End While}\\
\>\>  Assign pools in $\cP'$ to tags on array $k$\\ 
\>\>  $\cP \leftarrow \cP \setminus \cP'$\\ 
{\tt End While}
\end{tabbing}
\end{minipage}}
}
\caption{\label{fig.primer-del} The iterative primer deletion algorithm.}
\compact
\compact
\compact
\end{figure}

%%%%%%%%%%%%%%%%%%%%%%%%%%%%%%%%%%%%%%%%%%%%%%%%%%%%%%%%%%
\section{Experimental Results} 
\label{sec.results}

\begin{table}[t]
{\footnotesize
\begin{center}
\caption{\label{table.tags}
Tag Sets Selected by the Greedy Algorithm.}
\begin{tabular}{| c | c | c || c  c | c c || c c | c c |}
\hline
$l$ & $h_{min}/$ & $c$ &   
\multicolumn{4}{c||}{(C1)+(C2)} &
\multicolumn{4}{c|}{(C1)+(C2)+(C3)}
\\
\cline{4-11}
& $h_{max}$ & & 
tags & Bound & $c$-tokens & Bound & 
tags & Bound & $c$-tokens & Bound 
\\
\hline
\hline
  &  &8 &213 &275 &2976 &3584 &107 &132 &1480 &1726
\\
20 & --/-- &9 &600 &816 &7931 &9792 &300 &389 &3939 &4672
\\
 &  &10 &1667 &2432 &20771 &26752 &844 &1161 &10411 &12780
\\
\hline
\hline
  &  &8 &175 &224 &2918 &3584 &90 &109 &1489 &1726
\\
--  &28/32 &9 &531 &644 &8431 &9792 &263 &312 &4158 &4672
\\
  & &10 &1428 &1854 &21707 &26752 &714 &896 &10837 &12780
\\
\hline
\hline
 & &8 &108 &224 &1548 &3584 &51 &109 &703 &1726
\\
20 &28/32 &9 &333 &644 &4566 &9792 &164 &312 &2185 &4672
\\
 & &10 &851 &1854 &11141 &26752 &447 &896 &5698 &12780
\\
\hline
\end{tabular}
\end{center}
}
\compact
\compact
\compact
\end{table}

\compact
\paragraph{\bf Tag Set Selection.}
The greedy tag set design algorithm described in 
Section \ref{sec.tag-set-design} can be used 
to fully or selectively enforce the constraints in Definition 
\ref{def.feasible-tag-set}.
In order to assess the effect of various hybridization constraints on 
tag set size, we ran the algorithm both with constraints (C1)+(C2) and 
with constraints (C1)+(C2)+(C3).
For each set of constraints, 
we ran the algorithm with $c$ between 8 and 10
for typical practical requirements \cite{affy_note,affy_patent} that 
all tags have length 20 and weight between 28 and 32 
(corresponding to a \texttt{GC}-content between 40-60\%).
We also ran the algorithm with the tag length and weight 
requirements enforced individually.

Table \ref{table.tags} gives the size of the 
tag set found by the greedy algorithm, as well as the number of 
$c$-tokens appearing in selected tags.  We also include 
the theoretical upper-bounds on these two quantities; 
the upper-bounds for (C1)+(C2) follow from results of \cite{Karp00}, 
while the upper-bounds for (C1)+(C2)+(C3) follow from 
Lemma \ref{lemma.count} and Theorem \ref{thm.upperb}.
The results show that, for any combination of length and 
weight requirements, imposing the antitag-to-antitag hybridization 
constraints (C3) roughly halves the number of tags selected 
by the greedy algorithm -- as well as the theoretical upperbound -- 
compared to only imposing antitag-to-tag hybridization constraints (C1)+(C2). 
For a fixed set of hybridization constraints, the largest tag sets 
are found by the greedy algorithm when only the length 
requirement is imposed.  The tag weight requirement, which 
guarantees similar melting temperatures for the tags, 
results in a 10-20\% reduction in the number of tags. However, 
requiring that the tags have {\em both} equal length and 
similar weight results in close to halving the number of tags. 
This strongly suggests reassessing the need for the strict simultaneous 
enforcement of the two constraints in current industry 
designs \cite{affy_note}; our results indicate that 
allowing small variations in tag length and/or weight 
results in significant increases in the number of tags.

\compact
\paragraph{\bf Integrated Primer Selection and Tag Assignment.}
We have implemented the iterative primer deletion algorithm 
in Figure \ref{fig.primer-del} (Primer-Del), 
a variant of it in which primers in pools of size 1 are 
omitted -- unless all pools have size 1 -- when selecting 
the primer with maximum potential for deletion (Primer-Del+), 
and two simple heuristics that first select from each pool 
the primer of minimum potential (Min-Pot), 
respectively minimum degree (Min-Deg), 
and then run the iterative primer deletion algorithm 
on the resulting pools of size 1.  
We ran all algorithms on data sets with between 
1000 to 5000 pools of up to 5 randomly generated primers. 
As in \cite{BenDor03}, we varied the number of tags between 
500 and 2000.  

For instance size, we report the number 
of arrays and the average tag utilization (computed over 
all arrays except the last) obtained by (a) 
algorithm B in \cite{BenDor03} run using a single primer 
per pool, (b) the four pool-aware assignment algorithms 
run with 1 additional candidate in each pool, and 
(c) the four pool-aware assignment algorithms 
run with 4 additional candidates in each pool.  
Scenario (b) models SNP genotyping applications in which the primer 
can be selected from both strands of the template DNA, while 
scenario (c) models applications such as gene transcription 
monitoring, where significantly more than 2 gene specific primers are typically available.

In a first set of experiments we extracted tag sequences from the tag 
set of the commercially available GenFlex Tag Arrays.  All GenFlex 
tags have length 20; primers used in our experiments 
are 20 bases long as well. 
Primer-to-tag hybridizations were assumed to occur 
between primers and tags containing complementary 
$c$-tokens with $c=7$ (Table \ref{table.multiplexing7}), respectively 
$c=8$ (Table \ref{table.multiplexing8}).
The results show that 
significant improvements in multiplexing rate -- and 
a corresponding reduction in the number of arrays -- 
are achieved by the pool-aware algorithms over 
the algorithm in \cite{BenDor03}. 
For example, assaying 5000 reactions on a 2000-tag array 
requires 18 arrays using the method in \cite{BenDor03} for $c=7$, 
compared to only 13 (respectively 9) if 2 (respectively 5) 
primers per pool are available.
In these experiments, the Primer-Del+ algorithm dominates 
in solution quality the Primer-Del, while Min-Deg dominates 
Min-Pot.  Neither Primer-Del+ nor Min-Deg consistently 
outperforms the other over the whole range of parameters, 
which suggests that a good practical meta-heuristic is to 
run both of them and pick the best solution obtained.

In a second set of experiments we compared two sets of 
213 tags of length 20, one constructed by running the greedy 
algorithm in Section \ref{sec.tag-set-design} with $c=8$ and constraints 
(C1)+(C2), and the other extracted from the GenFlex Tag Array.  
The results in Table \ref{table.multiplexing213} show that 
the tags selected by the greedy algorithm participate in fewer 
primer-to-tag hybridizations, which leads to an improved 
multiplexing rate.

\begin{table}[t]
{\footnotesize
\begin{center}
\caption{\label{table.multiplexing7}
Multiplexing results for $c=7$ (averages over 10 test cases).}
\begin{tabular}{| c | c | c | c c | c c | c c |}
\hline
\# & Pool &  Algorithm &                 
\multicolumn{2}{c|}{500 tags} &
\multicolumn{2}{c|}{1000 tags} &   
\multicolumn{2}{c|}{2000 tags}    
\\
pools & size &   &       
\#arrays & \% Util. & 
\#arrays & \% Util. & 
\#arrays & \% Util. \\
\hline
\hline
 &1 &\cite{BenDor03}			 &7.5 &30.1	 &6.0 &19.3	 &5.0 &12.1	
\\
\cline{2-9}
 &2 &Primer-Del			 &6.0 &38.7	 &5.0 &24.3	 &4.1 &15.5	
\\
 &2 &Primer-Del+		 &6.0 &39.6	 &4.5 &27.3	 &4.0 &16.5	
\\
 &2 &Min-Pot		 &6.0 &38.4	 &5.0 &24.2	 &4.0 &15.9	
\\
1000 &2 &Min-Deg		 &5.8 &40.9	 &4.6 &27.0	 &4.0 &16.4	
\\
\cline{2-9}
 &5 &Primer-Del			 &5.0 &49.6	 &4.0 &32.5	 &3.3 &21.0	
\\
 &5 &Primer-Del+		 &4.0 &60.4	 &3.0 &43.6	 &3.0 &24.7	
\\
 &5 &Min-Pot		 &4.9 &50.6	 &4.0 &33.0	 &3.0 &23.5	
\\
 &5 &Min-Deg		 &4.0 &62.0	 &3.0 &44.9	 &2.7 &28.1	
\\
\hline
\hline
 &1 &\cite{BenDor03}			 &13.4 &31.8	 &11.0 &19.9	 &8.7 &12.9	
\\
\cline{2-9}
 &2 &Primer-Del			 &10.7 &41.0	 &8.5 &26.4	 &7.0 &16.6	
\\
 &2 &Primer-Del+		 &10.0 &43.3	 &8.0 &28.1	 &6.0 &19.1	
\\
 &2 &Min-Pot		 &11.0 &39.4	 &9.0 &24.8	 &7.0 &16.3	
\\
2000 &2 &Min-Deg		 &10.0 &43.5	 &8.0 &28.2	 &6.0 &19.2	
\\
\cline{2-9}
 &5 &Primer-Del			 &8.0 &56.8	 &6.1 &38.4	 &5.0 &24.5	
\\
 &5 &Primer-Del+		 &7.1 &62.4	 &6.0 &39.7	 &4.0 &30.1	
\\
 &5 &Min-Pot		 &9.2 &47.5	 &7.0 &32.9	 &5.0 &24.0	
\\
 &5 &Min-Deg		 &7.0 &63.1	 &5.3 &44.2	 &4.0 &30.7	
\\
\hline
\hline
 &1 &\cite{BenDor03}			 &29.5 &35.0	 &23.0 &22.6	 &18.0 &14.6	
\\
\cline{2-9}
 &2 &Primer-Del			 &22.2 &47.0	 &17.1 &30.9	 &13.7 &19.6	
\\
 &2 &Primer-Del+		 &22.2 &46.8	 &17.0 &30.9	 &13.1 &20.4	
\\
 &2 &Min-Pot		 &25.0 &41.5	 &19.2 &27.3	 &15.0 &17.7	
\\
5000 &2 &Min-Deg		 &22.0 &47.3	 &17.0 &31.0	 &13.0 &20.6	
\\
\cline{2-9}
 &5 &Primer-Del			 &16.6 &63.8	 &12.3 &43.9	 &10.0 &27.8	
\\
 &5 &Primer-Del+		 &16.0 &65.6	 &12.0 &44.9	 &9.0 &30.6	
\\
 &5 &Min-Pot		 &29.5 &35.0	 &23.0 &22.6	 &18.0 &14.6	
\\
 &5 &Min-Deg		 &16.0 &65.8	 &12.0 &45.2	 &9.0 &30.8	
\\
\hline
\end{tabular}
\end{center}
}
\end{table}

\compact
\compact
\compact

\begin{table}[t]
{\footnotesize
\begin{center}
\caption{\label{table.multiplexing8}
Multiplexing results for $c=8$ (averages over 10 test cases).}
\begin{tabular}{| c | c | c | c c | c c | c c |}
\hline
\# & Pool &  Algorithm &                 
\multicolumn{2}{c|}{500 tags} &
\multicolumn{2}{c|}{1000 tags} &   
\multicolumn{2}{c|}{2000 tags}    
\\
pools & size &   &       
\#arrays & \% Util. & 
\#arrays & \% Util. & 
\#arrays & \% Util. \\
\hline
\hline
 &1 &\cite{BenDor03}			 &3.0 &86.0	 &2.0 &77.1	 &2.0 &46.3	
\\
\cline{2-9}
 &2 &Primer-Del			 &3.0 &90.1	 &2.0 &81.6	 &2.0 &47.8	
\\
 &2 &Primer-Del+		 &3.0 &94.5	 &2.0 &88.5	 &1.0 &50.0	
\\
 &2 &Min-Pot		 &3.0 &94.4	 &2.0 &87.9	 &1.0 &50.0	
\\
1000 &2 &Min-Deg		 &3.0 &92.6	 &2.0 &88.8	 &1.0 &50.0	
\\
\cline{2-9}
 &5 &Primer-Del			 &3.0 &98.0	 &2.0 &92.6	 &2.0 &49.2	
\\
 &5 &Primer-Del+		 &3.0 &99.5	 &2.0 &97.4	 &1.0 &50.0	
\\
 &5 &Min-Pot		 &3.0 &99.4	 &2.0 &97.1	 &1.0 &50.0	
\\
 &5 &Min-Deg		 &3.0 &93.4	 &2.0 &93.4	 &1.0 &50.0	
\\
\hline
\hline
 &1 &\cite{BenDor03}			 &6.0 &78.2	 &4.0 &64.4	 &3.0 &48.3	
\\
\cline{2-9}
 &2 &Primer-Del			 &5.0 &92.3	 &4.0 &66.6	 &3.0 &49.8	
\\
 &2 &Primer-Del+		 &5.0 &93.5	 &3.0 &87.9	 &2.0 &78.7	
\\
 &2 &Min-Pot		 &5.0 &93.6	 &3.0 &87.7	 &2.0 &78.1	
\\
2000 &2 &Min-Deg		 &5.0 &90.8	 &3.0 &87.5	 &2.0 &79.6	
\\
\cline{2-9}
 &5 &Primer-Del			 &5.0 &98.4	 &3.0 &94.1	 &2.0 &84.8	
\\
 &5 &Primer-Del+		 &5.0 &99.5	 &3.0 &97.1	 &2.0 &91.2	
\\
 &5 &Min-Pot		 &5.0 &99.5	 &3.0 &97.0	 &2.0 &90.8	
\\
 &5 &Min-Deg		 &5.0 &91.8	 &3.0 &90.6	 &2.0 &91.7	
\\
\hline
\hline
 &1 &\cite{BenDor03}			 &13.0 &81.3	 &8.6 &64.7	 &6.0 &49.3	
\\
\cline{2-9}
 &2 &Primer-Del			 &12.0 &90.5	 &7.0 &81.1	 &5.0 &61.7	
\\
 &2 &Primer-Del+		 &11.2 &93.8	 &7.0 &81.9	 &4.0 &73.8	
\\
 &2 &Min-Pot		 &12.0 &90.4	 &7.0 &81.2	 &5.0 &62.2	
\\
5000 &2 &Min-Deg		 &12.0 &90.1	 &7.0 &81.5	 &4.0 &73.9	
\\
\cline{2-9}
 &5 &Primer-Del			 &11.0 &98.9	 &6.0 &96.1	 &4.0 &81.7	
\\
 &5 &Primer-Del+		 &11.0 &99.4	 &6.0 &96.8	 &3.0 &97.1	
\\
 &5 &Min-Pot		 &11.0 &99.4	 &6.0 &96.9	 &4.0 &83.1	
\\
 &5 &Min-Deg		 &11.0 &94.6	 &6.0 &91.0	 &3.4 &88.0	
\\
\hline
\end{tabular}
\end{center}
}
\compact
\compact
\compact
\end{table}

\begin{table}[t]
{\footnotesize
\begin{center}
\caption{\label{table.multiplexing213}
Multiplexing results (averages over 10 test cases) 
for two sets of 213 tags of length 20, one constructed by running the greedy             
algorithm in Section \ref{sec.tag-set-design} with $c=8$ and constraints
(C1)+(C2), and the other extracted from the GenFlex Tag Array.}
\begin{tabular}{| c | c | c | c c | c c | }
\hline
\# & Pool &  Algorithm &                 
\multicolumn{2}{c|}{GenFlex tags} &
\multicolumn{2}{c|}{Greedy tags} 
\\
pools & size &   &       
\#arrays & \% Util. & 
\#arrays & \% Util. \\
\hline
\hline
 &1 &\cite{BenDor03}                       &6.0 &90.0      &5.0 &100.0
\\
\cline{2-7}
 &2 &Primer-Del+      &5.0 &100.0     &5.0 &100.0
\\
1000 &2 &Min-Deg                 &5.9 &94.0      &5.0 &100.0
\\
\cline{2-7}
 &5 &Primer-Del+      &5.0 &100.0     &5.0 &100.0
\\
 &5 &Min-Deg                 &5.2 &97.3      &5.0 &100.0
\\
\hline
 &1 &\cite{BenDor03}                       &11.0 &90.6     &10.0 &99.2
\\
\cline{2-7}
 &2 &Primer-Del+      &10.0 &98.7     &10.0 &100.0
\\
2000 &2 &Min-Deg                 &10.8 &94.2     &10.0 &99.3
\\
\cline{2-7}
 &5 &Primer-Del+      &10.0 &100.0    &10.0 &100.0
\\
 &5 &Min-Deg                 &10.1 &96.0     &10.0 &99.3
\\
\hline
 &1 &\cite{BenDor03}                       &26.5 &91.3     &24.0 &99.2
\\
\cline{2-7}
 &2 &Primer-Del+      &25.0 &97.6     &24.0 &100.0
\\
5000 &2 &Min-Deg                 &25.0 &96.3     &24.0 &99.3
\\
\cline{2-7}
 &5 &Primer-Del+      &24.0 &100.0    &24.0 &100.0
\\
 &5 &Min-Deg                 &25.0 &96.6     &24.0 &99.3
\\
\hline
\end{tabular}
\end{center}
}
\compact
\compact
\compact
\end{table}

%%%%%%%%%%%%%%%%%%%%%%%%%%%%%%%%%%%%%%%%%%%%%%%%%%%%%%%%%%
%\section{Conclusions} 
%\label{sec:conclusions}

%%%%%%%%%%%%%%%%%%%%%%%%%%%%%%%%%%%%%%%%%%%%%%%%%%%%%%%%%%

\compact
{\footnotesize
%\small                 
\bibliography{longnames,microarray,myconf,primer}      
\bibliographystyle{plain}
}

\clearpage 
\appendix 

\section{Proof of Lemma \ref{lemma.count}}
\label{sec.appendix}

We first establish two lemmas on self-complementary DNA strings, i.e., 
strings 
$x\in\{\tA,\tC,\tT,\tG\}^{+}$
with $x=\overline{x}$. 

\begin{lemma}\label{lemma.even}
If $x$ is self-complementary then $|x|$ and $w(x)$ are both even.
\end{lemma}
\begin{proof}
Let $x=x_{1}x_{2}\ldots{x_{p}}$ be a self-complementary DNA string. 
If $p=2q+1$, by the definition of the complement we should 
have $x_{q+1}=\overline{x}_{q+1}$, which is impossible.
Thus, $p=2q$. Since 
$x_{1}=\overline{x}_{2q}$,$x_{2}=\overline{x}_{2q-1}$,$\ldots$,
$x_{q}=\overline{x}_{q+1}$, and the weight of complementary bases 
is the same, it follows that $w(x)=2\sum_{i=1}^{q}w(x_{i})$.
\qed
\end{proof}

\begin{lemma}\label{lemma.H_n}
Let $H_{n}$ be the number of self-complementary DNA strings of weight $n$.
$H_n=0$ if $n$ is odd, and $H_n=G_{n/2}$ if $n$ is even.
\end{lemma}
\begin{proof}
By Lemma \ref{lemma.even}, self-complementary strings must have even 
length and weight. For even $n$, the mapping 
$x_{1}\ldots x_{q}x_{q+1}\ldots x_{2q} \mapsto x_{1}\ldots x_{q}$ 
gives a one-to-one correspondence between self-complementary strings of 
weight $n$ and strings of weight $n/2$.
\qed
\end{proof}

%\begin{proof}
\noindent 
{\em Proof of Lemma \ref{lemma.count}.}
Let $\tW$ and $\tS$ denote weak and strong DNA bases ($\tA$ or $\tT$, 
respectively $\tG$ or $\tC$), and let 
$<$$w$$>$ denote the set of DNA strings with weight $w$.
The $c$-tokens can be partitioned into the seven classes 
given in Table \ref{table.classes}, 
depending on total token 
weight ($c$ or $c+1$) and the type of starting and ending bases. 
This partitioning is defined so that, for every $c$-token
$x$, the class of the unique $c$-token suffix of 
$\overline{x}$ can be determined from the class 
of $x$. 
Note that $\bar{x}$ is itself a $c$-token, except 
when $x\in \SciiiWW ~\cup~ \ScivSW$.

\begin{table}[b]
\caption{\label{table.classes} Classes of $c$-tokens.}
\begin{center}
\begin{tabular}{|c|c|}
\hline
Class of $x$ &  $c$-token suffix of $\overline{x}$\\
\hline
$\WciiiS$ & $\SciiiW$ \\
$\ScivS$ & $\ScivS$ \\
$\SciiiS$ & $\SciiiS$ \\
$\WciiW$ & $\WciiW$ \\
$\SciiiW$ & $\WciiiS$ \\
$\SciiiWW$ & $\WciiiS$ \\
$\ScivSW$ & $\ScivS$ \\
\hline
\end{tabular}
\end{center}          
\end{table} 

\noindent
Let $N_{\mathit{cls}}$ denote the number of $c$-tokens 
of class $\mathit{cls}$ occurring in a feasible tag set. 

\subsection*{$c$ odd}

Since $\WciiiS ~\cup~ \SciiiW$ 
can be partitioned into $4G_{c-3}$ pairs $\{x,\bar{x}\}$
of complementary $c$-tokens, and at most one token from 
each pair can appear in a feasible tag set, 
\begin{equation}\label{eq1}
	N_{\sWciiiS} + N_{\sSciiiW} \le 4G_{c-3}
\end{equation}
Similarly, class $\WciiW$ can be partitioned into 
$2G_{c-2}$ pairs $\{x,\bar{x}\}$ of complementary $c$-tokens, 
$\WciiiS ~\cup~ \SciiiWW$ can be partitioned 
into $4G_{c-3}$ triples $\{x,\bar{x}A,\bar{x}T\}$ with $x\in \WciiiS$, 
$\SciiiW ~\cup~ \SciiiWW$ can be partitioned 
into $4G_{c-3}$ triples $\{x,xA,xT\}$ with $x\in \SciiiW$, 
and 
$\ScivS ~\cup~\ScivSW$ can be partitioned 
into $2G_{c-4}$ 6-tuples $\{x,\bar{x},xA,xT,\bar{x}A,\bar{x}T\}$ with $x\in \ScivS$.
Since at most one $c$-token can appear in a feasible tag set from each 
such pair, triple, respectively 6-tuple,
\begin{equation}\label{eq2}
	N_{\sWciiW} \le 2G_{c-2}
\end{equation}
\begin{equation}\label{eq3}
	N_{\sWciiiS} + N_{\sSciiiWW} \le 4G_{c-3}
\end{equation}
\begin{equation}\label{eq4}
	N_{\sSciiiW} + N_{\sSciiiWW} \le 4G_{c-3}
\end{equation}
\begin{equation}\label{eq5}
	N_{\sScivS} + N_{\sScivSW} \le 2G_{c-4}
\end{equation}
Using Lemma \ref{lemma.H_n}, it follows that $\SciiiS$ 
contains $2G_{\frac{c-3}{2}}$ self-complementary $c$-tokens.
Since the remaining $4G_{c-3}-2G_{\frac{c-3}{2}}$ $c$-tokens
can be partitioned into complementary pairs each contributing 
at most one $c$-token to a feasible tag set, 
\begin{equation}\label{eq6}
	  N_{\sSciiiS} 
        \le 
	  \frac{1}{2}\left(4G_{c-3}-2G_{\frac{c-3}{2}}\right) + 2G_{\frac{c-3}{2}}
	= 
          2G_{c-3} + G_{\frac{c-3}{2}}
\end{equation}
Adding inequalities (\ref{eq1}), (\ref{eq3}), and (\ref{eq4}) multiplied by 1/2 
with (\ref{eq2}), (\ref{eq5}), and (\ref{eq6}) implies that the 
total number of $c$-tokens in a feasible tag set is at most 
\[
2 G_{c-2} + 8 G_{c-3} + 2 G_{c-4} + G_{\frac{c-3}{2}} 
= 3 G_{c-2} + 6 G_{c-3} + G_{\frac{c-3}{2}}
\]
Furthermore, adding (\ref{eq1}), (\ref{eq2}), and (\ref{eq3}) 
with inequalities (\ref{eq5}) and (\ref{eq6}) multiplied by 2 implies that 
the total tail weight of the $c$-tokens in a feasible tag set is at most 
\[
   2 G_{c-2} + 12 G_{c-3} + 4 G_{c-4} + 2 G_{\frac{c-3}{2}}
= 2 G_{c-1} + 4 G_{c-3} + 2 G_{\frac{c-3}{2}}
\]

\subsection*{$c$ even}

Inequalities (\ref{eq1}), (\ref{eq3}), and (\ref{eq4}) continue to hold for 
even values of $c$. 
Since $c-3$ is odd,  $\SciiiS$ contains no 
self-complementary tokens and can be partitioned 
into $2 G_{c-3}$ pairs $\{x,\bar{x}\}$,  
\begin{equation} \label{eq7}
	N_{\sSciiiS}
	\le 2 G_{c-3}
\end{equation}
By Lemma \ref{lemma.H_n}, there are $2 G_{\frac{c-4}{2}}$ self-complementary 
tokens in $\ScivS$. Therefore 
$\ScivS ~\cup~\ScivSW$ can be partitioned
into $2 G_{\frac{c-4}{2}}$ triples $\{x,xA,xT\}$ with $x\in \ScivS$, 
$x=\bar{x}$ and $2G_{c-4} - G_{\frac{c-4}{2}}$ 
6-tuples $\{x,\bar{x},xA,xT,\bar{x}A,\bar{x}T\}$ with 
$x\in \ScivS$, $x\neq \bar{x}$. Since 
a feasible tag set can use at most one $c$-token from each 
triple and 6-tuple, 
\begin{equation}\label{eq8}
	N_{\sScivS} + N_{\sScivSW} \le 2G_{c-4} + G_{\frac{c-4}{2}}
\end{equation}
Using again Lemma \ref{lemma.H_n}, we get 
\begin{equation} \label{eq9}
	N_{\sWciiW} \le 2 G_{c-2} + G_{\frac{c-2}{2}}
\end{equation}
Adding inequalities (\ref{eq1}), (\ref{eq3}), and (\ref{eq4}) multiplied by 1/2 
with (\ref{eq7}), (\ref{eq8}), and (\ref{eq9}) implies that the 
total number of $c$-tokens in a feasible tag set is at most 
\[
2 G_{c-2} + 8 G_{c-3} + 2 G_{c-4} + G_{\frac{c-2}{2}} + G_{\frac{c-4}{2}} 
= 3 G_{c-2} + 6 G_{c-3} + \frac{1}{2}G_{\frac{c}{2}}
\]
Finally, adding (\ref{eq1}), (\ref{eq3}), and (\ref{eq9}) 
with inequalities (\ref{eq7}) and (\ref{eq8}) multiplied by 2 implies that 
the total tail weight of the $c$-tokens in a feasible tag set is at most 
\[
   2 G_{c-2} + 12 G_{c-3} + 4 G_{c-4} + G_{\frac{c-2}{2}} + 2 G_{\frac{c-4}{2}}
= 2 G_{c-1} + 4 G_{c-3} + G_{\frac{c-2}{2}} + 2 G_{\frac{c-4}{2}}
\]
\qed
%\end{proof}

\end{document}